# The Reliability and Acceptance of Biometric System in Bangladesh: Users Perspective

Shaykh Siddique[*1], Monica Yasmin[*2], Tasnova Bintee Taher[*3], Mushfiqul Alam[*4]

[*]Department of Computer Science and Engineering, East West University, Dhaka, Bangladesh

*Abstract* — Biometric systems are the latest technologies of unique identification. People all over the world prefer to use this unique identification technology for their authentication security. The goal of this research is to evaluate the biometric systems based on system reliability and user satisfaction. As technology fully depends on personal data, so in terms of the quality and reliability of biometric systems, user satisfaction is a principal factor. To walk with the digital era, it is extremely important to assess users' concerns about data security as the systems are conducted the authentication by analyzing users' personal data. The study shows that users are satisfied by using biometric systems rather than other security systems. Besides, hardware failure is a big issue faced by biometric systems users. Finally, a matrix is generated to compare the performance of popular biometric systems from the users' opinions. As system reliability and user satisfaction are the focused issue of this research, biometric service providers can use these phenomena to find what aspect of improvement they need for their services. Also, this study can be a great visualizer for Bangladeshi users, so that they can easily realize which biometric system they have to choose.

## I. INTRODUCTION

With the development of technology, personal data security and privacy is considered a major challenge for all over the world. Biometrics systems are used to protect data and identity verification [1]. The safety issues of biometric information cannot be compromised. Where the issues have appeared, it may affect the quality and users' reliability. Some published research papers are related to biometric information privacy [2] and few define the critical issues of biometric information security [3]. A good reliability rate can increase biometric system users rather than using passwords or pin codes and poor reliability will reduce users. The users' trustworthiness, as well as the system security, are defined by using the "reliability" keyword here.

Online banking service [4], employee attendance, employee time tracking device and other identity verification systems are frequently used in Bangladesh. The most common methods of verification fingerprint and face recognition. On 2nd October 2016 Bangladesh government launched the Smart National ID card [5] and started collecting biometric information from the citizens. Recently to digitalize government services, the NID verification gateway server is getting under way [6]. With the large possibility of using biometric technology, it is very important to study the reliability and the trustworthiness of all biometric systems which are used in Bangladesh from users' point of view. Most of the biometric systems are integrated with hardware and software. The biometric system market will increase by 15% compound annual growth rate between 2017 and 2023 [7]. The market demand for new biometric systems can be visualized by analyzing adoption, satisfactoriness and reliability reviews.

The aim of the research is to evaluate the biometric systems in terms of system reliability and user satisfaction. Objectives are:

1) To analyze user reviews about the quality and reliability of biometric systems.
2) To assess user concerns about biometric data privacy.
3) To compare different biometric systems based on users' perspectives.

The paper has been organized into six sections, and after summing up those sections the full study can be visualized easily. Section I introduces biometric systems and their major challenges towards life. Some research objectives and questions are included so that after answering those questions and resolute the objectives, the aim of this paper can be fulfilled. Limitations of this study have also been described in this portion. Portion II collaborates the study with previous research papers and describes the above papers methodologies and founded results. The procedures which will be applied in this study and gives a summarized idea about data collection have been described in section III. Results have been extracted by collected data and described briefly in section IV, where Section V conducted with the discussion portion which describes the overall study and the effectiveness of this research. Section VI increases the consciousness of newcomers for the elaborate research about Biometric System as it represents the future work. The study has been concluded in section VII, and after that reference has been provided to verify the important pieces of information that make this study more effective.

The main study question is, what are the performances of biometric systems according to system reliability and user satisfaction? Also, have some sub-questions:

1) What are the users' opinions about the quality and reliability of Biometric Systems?
2) What is the user's level of concern for biometric data privacy?
3) What are the comparisons of biometric systems in terms of users' perspective?

The study was conducted on 175 people with an age limit of 18-50 years. Because of resource limitations, the survey questions were supplied using Google form through the internet. As the data were gathered through the internet, so only the people who used the internet were our participants. As the study portion was not that much vast, so there will be a good chance of result variation for a mass number of general people. Also, many participants might be the new users with less knowledge about biometric systems or many participants might be aged and not that much familiar with the biometric systems. This might contradict the result. Besides, we provided Google form via the internet, so there is a strong chance that the participant himself did not answer the questions or who responded to the questions was not the accurate person to participate in the survey. As the survey was executed on the internet so we could not monitor the participants' age. This might affect the result, arising invalid output. The question understanding gap is another point for result inaccuracy. Also, participants might not be interested to continue the survey or they might be in a hurry. But the collected data are stored safely which can reproduce for further findings. To avoid the garbage data, we assure that there was no lack of sincerity, so the validity of result accuracy is ensured. In this study, we only focused on users' perspective and their satisfaction in terms of biometric systems reliability and measured how good the quality of different biometric systems have but we did not focus to resolve the problems faced by users while they are using the biometric systems. Also how to enrich the quality of imperfect biometric systems was not focused on our study.

## II. RELATED WORKS

The traditional way of assuring safety and security, password protocol systems are used. But this protocol has some drawbacks: it can be stolen, users can forget their password. As this security issue has become a talk of the topic all over the world, the UK government started using biometric traits for identification in 1960. After this several studies took place for designing a biometric system and having a good recognition accuracy [8]. This technology uses the physical or behavioral traits of the user to solve the problem of authentication. The biometric system is consisting of four modules which are sensor module, feature extraction module, template database and a matching module. The process of authentication goes through in two-stage: Enrollment stage and verification stage. When a user puts his thumb on the sensor a picture of the fingerprint is taken by the sensor module. Further from this picture the system extracted some data and make it suitable to generate template data and save them in a database for the verification stage. After that, a query is made for matching the data with the template database to make sure that this user is a valid person. Imposing this for solving the security issue there arises two concerns one is biometric traits cannot be revoked and reissued when the biometric information of a person is compromised. If a person's fingerprint image is stolen it cannot be replaced in the template database as this information is unique.

Ratha [9] has detected eight points from where this biometric information can be stolen or tempered. Namely, attacks on the interface, attacks on the modules: feature extraction module and matching module, attacks on the template database and also the channels from where the biometric information is flowing. The result shows that the acceptance rate of fake fingerprints 67%. Liveness detection is a well-known countermeasure to distinguish between the fake fingerprint and the real fingerprint. Two scientists proposed to detect the perspiration phenomenon to differentiate the living fingers from the fake non – living fingers. Another researcher Coli Er Al. utilized the static feature as well as the dynamic features of a fingerprint image which to prevent the spoofing attack. Galbally [10] proposed a method that uses fingerprint parameterization based on the quality of fingerprint images. Kim, a researcher proposed to design an image descriptor to handle liveness detection. He used a property of image which is dispersion. The difference of dispersion in the image gradient field will be different if the fingerprint is being faked. This accuracy of a biometric system can be measured by using three factors: False Accept Rate (FAR), False Reject Rate (FRR) and Equal Error Rate (EER). The accuracy depends on image quality and matching algorithms. There is a platform which is FVC-ongoing where the researcher can upload their matching algorithm to evaluate the FAR, FRR and ERR. At present, the best matching algorithm which name is HXKJ and EER = 0.022% [8].

Among 4.6 billion mobile users 52.7% browse the Internet which is increasing cybercrimes rapidly. Bangladesh Government initiated a biometric SIM registration program in December 2015. Aim of this program has to verify the real owner of the SIM using NID and help to unearth the real criminal. S. I. Ahmed demonstrates about the ownership, user identity, exploitation, also security and safety concerns that challenged this program by using the method interview. Their online survey concluded that 77% of participants dislike the SIM registration system. Only 15% supported and 4% ignored the survey [2]. Many countries like Norway, Sweden and the US follow the law to protect data from a legal perspective [11]. The main motto of the study is to discuss several controversial legal issues rises from biometric context and analytical opinions conduction from the view of legal perspective on biometrics and data protection. Previous history observation methodology is used to run the study. Marcos [12] defined the difficulty of a biometric system with a study. Three main authentication methods are handheld token (card, ID, passport, and others), knowledge-based (password, PIN, and more others). Biometric technologies like fingerprint, voice, iris, face reflects some vulnerabilities. To detect the vulnerability of the biometric system two situations are intensely focused. The first one is impostors try to access the system as a substitute for a real user and another one is when a person avoids his identity which is suspenseful. The real-life challenges and their linkage to social, economic and political era are still untouched in this study.

### III. METHODOLOGY

As the study focuses on user reliability and acceptance, the answer to these research questions must be solved by the user's perspective. User reliability and acceptance level can be analyzed with numeric values. So our selected data type was quantitative. To find out the final outcome some mathematical functions and data analysis tools were used.

#### A. Data Collection Methods

For this quantitative study, questionnaires are more applicable rather than an interview, observation, discussion or other methods. It collects public opinion and gives a suitable result that can help to visualize any problem easily. It is cost-effective as well as, it saves time. The user's aptitude, consciousness, and opinion is the main agenda here. If this study conduct with security issues and the data alignment work, then company opinion such as- Grameenphone customer service, NID issue office, has to be considered. But this study works with the reliability of users and how they feel about adopting a biometric system as security is the concern here. Although less data accuracy, missing data, meaningless response, etc. are some drawbacks, it's an acceptable method for the study. This is a common and efficient way to reach a large number of participants easily.

We used 'Internet Survey' for this study. It deals with public opinion and gives a suitable result that can help to visualize any problem easily. The online survey is comparatively time savior, free of cost and helps to gather a huge amount of targeted data. Many people in Bangladesh use biometric authentication systems through smartphones. This method is an effective research method because the whole process will be done using the internet. As the whole process becomes autonomous, it eliminates the human error factor [13]. For collecting the price of the biometric system we used online markets like Amazon, Alibaba, etc.

#### B. Participants / Sampling

The study focus on users of biometric systems. So the participates of our survey will be general people with random ages and gender who uses the biometric system for identity verification. All the mobile phone users are the total population. According to Bangladesh Telecommunication Regulatory Commission, there are 159.780 million mobile phone users at the end of March 2019 [14]. The required sample size was 175 with a 95% confidence level and the confidence interval was 8 between selections.

Who has a mobile phone SIM Card, must have registered with fingerprint [2] and government National ID number. Also, smartphone users are using different biometric systems like- fingerprint, face recognition, etc.

#### C. Data Analysis

This qualitative study converted the raw data in numerical form to generate the output in quantity. Using statistics, some analyses made those quantitative data organized. The data analysis followed a process which involved in five stages [13].

*a) Data Preparation:*

Data were coded, pre-processed, cleaned from the collected raw data. Data were categorized and checked the data by the use of coding [13]. This study took a survey from Google form, that made ease of scaled up the raw data which regarded pre-processing. In the survey, if any user did not use biometric systems, the rest of the survey under this question was discarded which was indicated the cleaning method of raw data.

*b) Initial Exploration:*

After the data had been prepared, the initial exploration looked up for correlations [13]. In terms of exploring the raw data, prepared data was clustered and correlated in this study. If there was no relation between prepared data, researchers of this study did not approach this.

*c) Analysis:*

The analyses were compacted with the explored data and gave generated results by statistical tests linking to research questions [13]. This qualitative study formatted data in numerical value and analyzed them quantitatively. Data were compared and described with statistical analysis. Each objectives questionnaire was extracted by the measurement of straightforward central tendency analysis [13]. When researchers found a separate central tendency, those were compacted in a single result using mean calculation. The score of each biometric system+ was generated with gaining ratings,

$$score_{(i)} = \frac{\sum_{j=1}^{n} point_j \times 100}{Total\ Score} \quad (1)$$

Equation (1), i for each parameter, j is for each record and Total Score = Total Participants × 5.

*d) Presentation:*

This study had been represented by different graphs, charts, and matrix. Statistically analyzed data was represented by a pie chart, as segments of the whole pie which was visually powerful [13]. As raw data was clustered and correlated, some plots were used to organize the presentation. These results were put into a two-dimensional matrix. Matrix mapped the user's measurement parameters with different biometric systems, which fulfilled the researcher's aim of this study.

*e) Tools:*

Researchers used a Python tool to generate the statistical analysis easy and presentation more equipped. Using Jupiter Anaconda, data were analyzed and represented in graphs and charts collaborated with MS Word.

*f) Research Ethics:*

As we collect users' personal opinions about a security system, so we must ensure data privacy. The data collection process was anonymous for participants. All the personal information was secured and will never be exposed.

## IV. RESULTS

The research study was conducted with the reliability and acceptance of biometric System in Bangladesh for the users' perspective. The data collection process of our study was a voluntary online survey. We have gathered 175 responses who wanted to participate in our survey voluntarily. The response rate was 98.85% of the total participants.

### A. User Reliability Reviews

Participants were asked to give their opinion about which system they prefer to use in between Biometric System and Password/pin based security system. We categorize their responses into five categories.

Fig. 1 shows that, here in Bangladesh 63.63% of people think that biometric systems can give more secure than password or pin base security system.10.32% of people totally disagreed that biometric System is not secure at all. 26.45% of people do not want to give their opinion on this question.

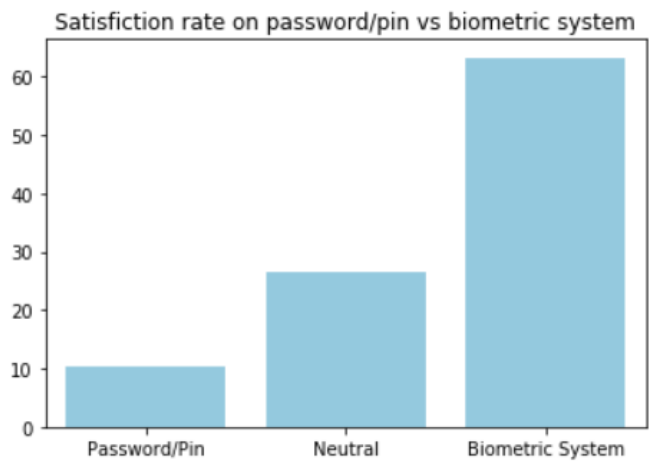

**Figure 1: Satisfaction rate of password/pin and biometric system**

For Fig. 2, participants were asked, what kind of issue they faced while using a biometric system. We gave our participants some options about which kind of problem they faced while using a biometric system.

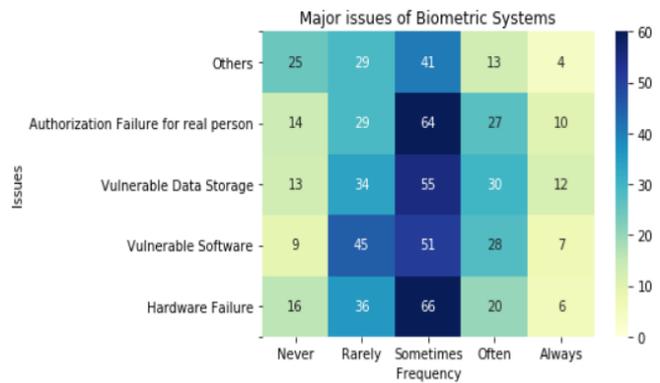

**Figure 2: Major issues of biometric systems**

Fig. 2 shows that among the participants, 66% score for those who faced hardware failure sometimes. One of the major issues is authorization failure for real person for any kind of biometric system. 64% score for authorization failure of the real person. 51% score of participants stated that they are using vulnerable software which might leak biometric information to the wrong hands. Vulnerable data storage scored 55%. The deep coloring part of the graph represents the scores for those particular issues. Scores were measured with Eq. 1.

### B. User Concerns About Biometric Data Privacy

From Fig. 3 pie chart shows awareness about fake authentication of biometric systems. Participants were asked about an imaginary scenario to extract their thinking capability to fool a biometric system. 36% of participants think that it is possible to fool a biometric system and 26% voted for "often" which is encoded in value "4", which also indicated the positive reaction of participants to fool a biometric system. 18% of people were strongly agreed with this and about 7% of the participants said it was not possible.

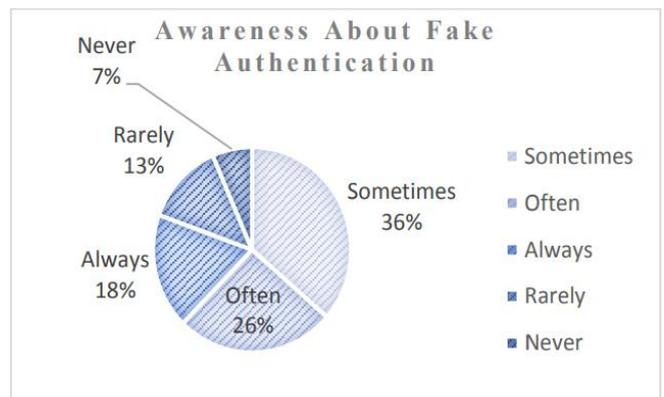

**Figure 3: Awareness about fake authentication**

Concerns about the recovery mechanisms and documentation/terms are shown in Fig. 4. Participants were asked about the reading of documentation or terms and conditions as well as if anything went wrong to anyone's skin or body which secondary recovery mechanism they used.

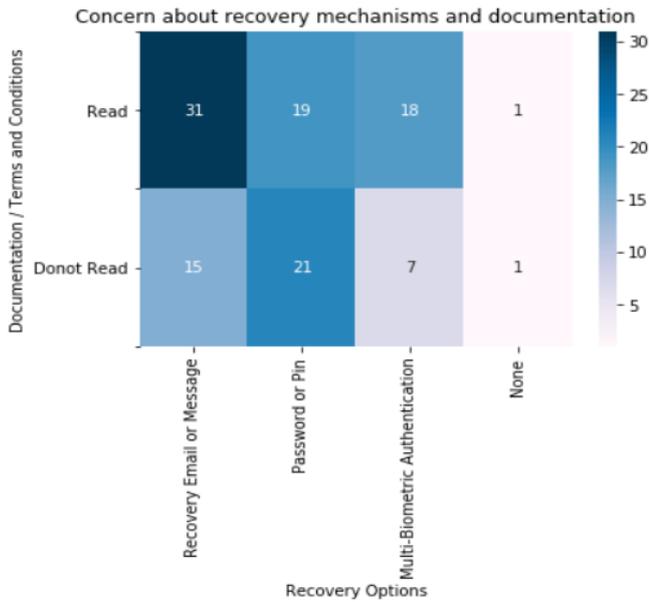

**Figure 4: Recovery mechanisms and documentation**

The rate of using a good recovery mechanism is high for the people who read the documentation papers. About 31 people chosen recovery email as a recovery mechanism and some people also chosen multi-biometric based authentication. A good range of people chosen password/pin as a secondary recovery mechanism. 15 participants did not read documentation and terms and conditions but chosen recovery email as a recovery mechanism. The percentage of documentation or terms and conditions the reader is higher and only 2 participants did not choose any recovery option.

*C. Comparisons of Biometric System*

The descriptive statistical analysis stated that the mean is above 3.35 when a user is using the biometric systems for their personal purpose. Whereas the shared biometric system has less than 3 which means participants have enough data privilege to use the personal biometric system rather than using a shared system according to Table 1.

**TABLE 1: PERSONAL AND SHARED DEVICE PRIVILEGES**

|  | Personal Device Privilege | Shared Data Storage privilege |
|---|---|---|
| Mean | 3.35 | 2.56 |
| Median | 4.00 | 2.00 |
| Std. Dev. | 0.99 | 0.94 |

The calculated median shows that personal devices have enough privilege to update, modify and remove biometric data, but shared data storage like National Biometric Data Storages, Telecommunication Company Biometric Data Storage did not have enough privilege.

Table 2 shows the score of some biometric systems according to the reliability, ease of use, security, cost and overall score. All perimeters were measured according to the score given by users. Equation (1) was used to evaluate the scores. The cost scores were measured by online market data. Overall performance depends on reliability, usability and security.

**TABLE 2: TWO DIMENSIONAL COMPARISON MATRIX: USER PERSPECTIVE**

| Types | Reliability | Ease of Use | Security | Cost | Overall |
|---|---|---|---|---|---|
| Face Detection | 46.15 | 52.70 | 52.45 | 12.05 | 50.43 |
| Fingerprint Scan | 58.24 | 64.78 | 60.63 | 10.54 | 61.21 |
| Hand Geometry | 48.81 | 46.29 | 47.42 | 16.74 | 47.51 |
| Iris Scan | 62.89 | 49.68 | 63.52 | 14.05 | 58.70 |
| DNA Test | 68.67 | 45.91 | 67.17 | 84.02 | 60.59 |
| Signature Validator | 44.02 | 43.14 | 43.52 | 9.36 | 43.56 |
| Voice Detector | 44.53 | 51.32 | 45.16 | 10.94 | 47.01 |

From Table 2, DNA test got the highest scores 68.67 and 67.17 in terms of reliability and security respectively. The Fingerprint scan got 64.78 score which defines the best usability. The cost of DNA Verification Device is very high, but fingerprints and signature validators are cheap. Fingerprint scored top in overall performance with 61.21 scores and then came DNA Test with 60.59 scores. The overall performance of Signature Validator was worst of all. For all cases, the scores were measured out of 100 and each score value represented the gaining score according to ratings.

## V. DISCUSSION

From the result analysis section, we are able to know that, among all the biometric systems, the fingerprint scans and the DNA verification systems are acceptable and reliable to the users. Top scored biometric systems specified the answer to the main research question of this study. DNA verification systems are secured, but on the other note, the fingerprint is tremendously easy to use. It affects the users' overall preference score with respect to other biometric systems. The result implies a research paper [15], where the researcher defined the best performing biometric system based on measuring accuracy, security and other comparatives.

According to the first research sub-question (What are the users' opinions about biometric systems based on quality and reliability?) users are now relaxing with biometric systems. Among the participants, 63.63% of the biometric systems users are satisfied with biometric authentication rather than the password or the pin-based authentication. People do not need to remember a strong password to secure their authentication as well as no risks of leaking passwords. The growth of collaborating biometric authentication with our personal life is increasing day by day. Besides, the issues faced by biometric users are also increasing. Sometimes, the users have to face hardware problems, vulnerable data storages and even authentication failure for real persons which compile with a piece of news [16] of the New York Times. The US embassy was stopped their visa immigration services for hardware failure of the biometric systems. About 50,000 visa applications were piled up.

More than 70% of the people said that sometimes fake authentication might be possible. Right documentation, terms and conditions, legal objectives can be used to improve the data privacy of biometric systems. The second research sub-question relates the public concerns with these kinds of biometric data privacy objects. If the system failed to authenticate, the recovery mechanisms are used to restore data. Our result figures out that, 60% of the participants from the survey read documentation. Among them, 15% preferred multi-biometric authentication. Now a smartphone device has many biometric authentication systems like fingerprint, face detection, voice detection at the same time. But 27% of the participants used email and message recovery options. If the recovery mechanism is not secure, then the strong biometric system's security has no value. Troy Hunt [17], Australia-based cybersecurity researchers dumped 700 million email addresses with weak passwords. So biometric data privacy becomes a cause of anxiety. Our study results covered these types of serious public concerns.

The users should have exact control over the access to biometric data. As claimed by the answer to the third research question, we had to do some comparisons of biometric systems. The result indicates that biometric systems users had enough privileges to update, modify and remove data from personal biometric devices, but they were not adequating with public biometric data storages (NID, VISA/passport, banking, telecommunication companies' data center). Except for the privileges, the biometric authentication system has some other comparative user-specific parameters like reliability, comfortability, security and cost. Our research result states that the DNA verification systems had the highest reliability and security from the users' perspective. Also, the findings of the results showed that fingerprint authentication was a highly comfortable device for biometric systems users. It saved time and money both. In the research paper [15], the researcher observed, tested and analyzed which confirmed our findings.

Our data collection was done through online. So the people who have the internet connection were our only participants and most of them were from the Dhaka region in Bangladesh. There is a strong possibility that some participants might be lost their interest while they were doing the survey or they were in a hurry. These kinds of situations might be affected and increased the deviation of our results. But we tried our best to dropout those kinds of responses.

## VI. FUTURE WORK

As this study worked with users' opinions and extracted new results by analyzing collected data, there can be a further study based on Biometric systems providers' perspective. Several companies such as - DERMALOG, ZKTeco, TigerIT provide Biometric Systems. The perception of company stakeholders about Biometric systems can be new work. Apart from the mentioned companies, government laws about Biometric systems can be another future study. This study focused on users' satisfaction over Biometric Systems reliability, so finding the solution of given users' reviews of this study can be a new research work for newcomers.

## VII. CONCLUSIONS

In recent years, biometric systems fabricate a strong position and throw challenges to traditional methods of person authentication in the technological field. Its users' adoption rate is increasing which impacted our society also. In this paper, we tried to figure out the reliability of biometric systems and tried to measure the level of satisfaction from the users' perspective. This research tells us that users prefer to use biometric systems but they face many issues while using the systems. We made an effort to find out some issues faced by the biometric systems users. As the biometric systems are widely using technology for identification worldwide, issues faced by the users should be solved immediately. Many researchers are trying to resolve those issues and some issues are already been solved by previous research. If our research can contribute to further research to solve any issues then our research will be effective. If our findings from the research are considered effectively, then better improvement of biometric systems can be confirmed and the users would rely on those systems more. Systems reliability would be increased based on the trust of users.

## REFERENCES

[1] R. E. O. Paderes, A Comparative Review of Biometric Security Systems, Proc. - 8th Int. Conf. Bio-Science Bio-Technology, BSBT 2015, 8–11, 2016.

[2] S. I. Ahmed, M. R. Hoque, S. Guha, M. R. Rifat, and N. Dell, Privacy, security, and surveillance in the global south: A study of biometric mobile SIM registration in Bangladesh, Conf. Hum. Factors Comput. Syst. - Proc., (2017) 906–918.

[3] V. Andronikou, D. S. Demetis, T. Varvarigou, I. Group, and H. Street, Biometric Implementations and the Implications for Security and Privacy, 1st in-house FIDIS J. (1)(2007) 1–20.

[4] Standard Chartered Bank - Touch Registration and Login. [Online]. Available: https://www.online-banking.standardchartered.com/login/IBank?ser=100&act=TouchIdRegistration.jsp&cntry=BD. [Accessed: 24-Sep-2019].

[5] Bangladesh Introduces 'Smart' National Identity Cards - Global Voices Advox. [Online]. Available: https://advox.globalvoices.org/2016/10/07/bangladesh-introduces-smart-national-identity-cards/. [Accessed: 24-Sep-2019].

[6] Joy launches NID verification gateway server | Dhaka Tribune. [Online]. Available: https://www.dhakatribune.com/bangladesh/government-affairs/2019/07/17/joy-to-launch-nid-verification-gateway. [Accessed: 24-Sep-2019].

[7] Global Biometric System Market Research Report- Forecast 2023 | MRFR. [Online]. Available: https://www.marketresearchfuture.com/reports/biometric-system-market-3754. [Accessed: 24-Sep-2019].

[8] W. Yang, S. Wang, J. Hu, G. Zheng, and C. Valli, Security and accuracy of fingerprint-based biometrics: A review, Symmetry

(Basel)., 11(2) (2019).
[9] N. K. Ratha, S. Chikkerur, J. H. Connell, and R. M. Bolle, Generating cancelable fingerprint templates, IEEE Trans. Pattern Anal. Mach. Intell., 29(4) (2007) 561–572.
[10] J. Galbally, F. Alonso-Fernandez, J. Fierrez, and J. Ortega-Garcia, A high performance fingerprint liveness detection method based on quality related features, Futur. Gener. Comput. Syst., 28(1) (2012) 311–321.
[11] Y. Liu, Identifying legal concerns in the biometric context, J. Int. Commer. Law Technol., 3(1) (2008) 45–54.
[12] M. Faúndez-Zanuy, On the vulnerability of biometric security systems, IEEE Aerosp. Electron. Syst. Mag., 19(6) (2004) 3–8.
[13] Martyn Denscombe, The Good Research Guide for small-scale social research projects, Open Univ. Press. McGraw-Hill Educ., vol. Fourth Edi, (2012) 7–200.
[14] Mobile Phone Subscribers in Bangladesh March, 2019. | BTRC. [Online]. Available: http://www.btrc.gov.bd/content/mobile-phone-subscribers-bangladesh-march-2019. [Accessed: 29-Oct-2019].
[15] P. Ambalakat, Security of Biometric Authentication Systems, 21st Comput. Sci. Semin., no. SA1-T1-1.
[16] Computer Failure Leaves State Dept. Unable to Issue Visas - The New York Times. [Online]. Available: https://www.nytimes.com/2015/06/23/world/americas/computer-failure-leaves-state-dept-unable-to-issue-visas.html?smid=tw-nytimes. [Accessed: 26-Nov-2019].
[17] Troy Hunt: The 773 Million Record 'Collection #1' Data Breach. [Online]. Available: https://www.troyhunt.com/the-773-million-record-collection-1-data-reach/. [Accessed: 26-Nov-2019].